\documentclass[twocolumn,amsmath,amssymb,10pt,superscriptaddress,a4paper,letterpaper,fleqn]{revtex4-2}
 \usepackage{amsmath}
 \usepackage{amssymb}
 \usepackage{epsfig}
 \usepackage{graphicx}
 \usepackage{dcolumn}
 \usepackage{array}
 \usepackage{bm}
                                        
 \usepackage{longtable}
 \usepackage{multirow}
 \usepackage{float}
                                      
 \usepackage{afterpage}
 \usepackage{color}
 \usepackage{footnote}
 \setlongtables
 \usepackage[breaklinks=true,linkbordercolor={1 1 1}]{hyperref}
 
\begin{document}
 \setcounter{page}{1}
 
 
 \title{Can decay heat measurements tell us something about the Reactor Antineutrino Anomaly?}
                                   
  \author{A.~A.~Sonzogni}
  \email{sonzogni@bnl.gov}                           
 \affiliation{Nuclear Science \& Technology Department, Brookhaven National Laboratory, Upton, NY 11973-5000, USA}

  \author{R.~J.~Lorek}
  \affiliation{National Nuclear Data Center, Brookhaven National Laboratory, Upton, NY 11973-5000, USA}

  \author{A.~Mattera}
  \affiliation{National Nuclear Data Center, Brookhaven National Laboratory, Upton, NY 11973-5000, USA}
  
 \author{E.~A.~McCutchan}
  \affiliation{National Nuclear Data Center, Brookhaven National Laboratory, Upton, NY 11973-5000, USA}

 \date{\today}
 
 \begin{abstract}
 {
Measurements of the decay energy released as a function of time following the thermal neutron induced fission of  $^{235}$U and $^{239,241}$Pu were performed in the 1970s at Oak Ridge National Laboratory with the purpose of quantifying possible Loss Of Coolant Accident scenarios.   The derivative of this decay energy with respect to time, known in technical parlance as decay heat, is mainly composed of two terms, that of the electrons produced together with antineutrinos in the beta-minus decay of the neutron-rich fission products, and that of the gamma rays produced in the subsequent decay of excited nuclear levels.  
In this work we study if this  extensive set of decay energy measurements can be used to assess the Reactor Antineutrino Anomaly, that is, the approximately 5\% deficit of electron antineutrinos produced by nuclear reactors, first deduced by Mention and collaborators in 2011, and observed by the major reactor antineutrino experiments since.  
With the assistance of nuclear databases, we are able to obtain the ratio of electron spectra under equilibrium conditions for $^{235}$U to $^{239}$Pu, in better agreement with the lower
trend recently reported by Kopeikin and collaborators, 
as well as those for $^{235}$U to $^{241}$Pu and $^{241}$Pu to $^{239}$Pu, which do not agree well with those measured at the Institut Laue-Langevin in the 1980s.
We conclude that a new experimental campaign is needed to measure the electron spectra utilizing a high-resolution and signal-to-noise-ratio electron spectrometer and a highly precise fission normalization procedure.
 }
 \end{abstract}
 \maketitle

\section{Introduction}

 We have witnessed remarkable developments in the field of nuclear reactor antineutrinos in the last 10 years,
 starting with the more precise spectra estimates of Huber~\cite{huber11} and Mueller {\it et al.}~\cite{mueller11}; as well as the detailed analysis
 of Mention {\it et al.}~\cite{mention11}, which led to the conclusion that approximately 5\% of the electron antineutrinos are missing
 at short distances, a feature that has been coined as the Reactor Antineutrino Anomaly (RAA).   
 On the experimental side, the Daya Bay~ \cite{dayabay16}, Double Chooz~\cite{doublechooz} and RENO~\cite{reno16} collaborations have measured the $\theta_{13}$ mixing angle with great precision;
they have also  revealed a deficit of antineutrinos at the peak of the Inverse Beta Decay (IBD) antineutrino spectrum and a small excess at around
 5 MeV with respect to the Huber-Mueller model, also observed by the short-distance  NEOS experiment~\cite{neos}. 
 More recently, the short-distance  PROSPECT~\cite{prospect19} and STEREO~\cite{stereo20} collaborations have  published their measured $^{235}$U spectra, while PROSPECT-Daya Bay~\cite{DBP} and PROSPECT-STEREO~\cite{PS} joint analyses were just published.
Finally, the results of the long-baseline JUNO experiment~\cite{juno} are eagerly anticipated by the community.
 
In order to precisely account for the electron antineutrinos produced by nuclear reactors, we need to have accurate predictions of the antineutrino spectra produced by each of the main
nuclides undergoing fission,  $^{235,238}$U and $^{239,241}$Pu.
Our current best numerical estimates of the $^{235}$U and $^{239,241}$Pu antineutrino spectra are obtained from a multi average-beta-decay-branch fit performed by P. Huber~\cite{huber11} to the corresponding electron  spectra measured at the Institut Laue-Langevin (ILL) in the 1980s~\cite{ill82, ill85, ill89}.
These electron spectra were measured using the BILL spectrometer, which provided excellent energy resolution and signal-to-noise ratio, from foils placed inside the ILL reactor.
For  $^{238}$U, the summation calculation from Mueller {\it et al}~\cite{mueller11} is currently considered its best antineutrino spectrum estimate.   
A number of hypothesis have been postulated to explain the disagreement between the latest reactor spectrum measurements with the Huber-Mueller model, 
including  forbidden beta-minus transitions effects~\cite{hayen19} and incomplete beta-minus decay schemes~\cite{letouneau23}, to name just a few. 
In this work we explore in detail some of the ILL measurements underpinnings to find a possible explanation of the RAA, including a comparison with data originally taken to quantify decay heat. 

\section{ILL Spectrum Normalization}

An early indication about possible issues in the $^{235}$U ILL data came from Daya Bay's measurement of the IBD antineutrino yield as function of the $^{239}$Pu fission fraction~\cite{dayabay17}, 
which concluded that faulty modeling was responsible for the RAA since their deduced $^{239}$Pu IBD yield was in agreement with Huber's value, while the $^{235}$U IBD yield was not.
The Daya Bay collaboration would later obtain $^{235}$U and $^{239}$Pu spectra, by themselves~\cite{dayabay19} and jointly with PROSPECT~\cite{DBP}, reaching similar conclusions.

For the normalization of the ILL electron spectra, that is, the derivation of absolute number of electrons at a given energy per unit energy per fission,
a precise value of the neutron flux inside the reactor was needed, 
which was obtained by measuring the intensity of selected conversion electrons.  
From the brief technical description in the ILL articles, we know that the $^{235}$U experiment employed conversion electrons following  
neutron capture on $^{115}$In and $^{207}$Pb; the $^{239}$Pu one following neutron capture on  $^{115}$In and $^{197}$Au; 
and the $^{241}$Pu one following neutron capture on $^{113}$Cd, $^{115}$In and $^{207}$Pb.
We have checked all the cross section and conversion coefficient values quoted in the ILL normalization procedure, concluding that they are fairly close to the currently accepted best values, 
with the exception of the  $^{207}$Pb thermal neutron capture cross section $\sigma_{n\gamma}$($^{207}$Pb).
The ILL group used a  $\sigma_{n\gamma}$($^{207}$Pb) value equal to 712$\pm$10 mb  from the 1981 cross section evaluation work of S.F.~Mubhaghab~\cite{mughabghab81}.   
This value originates from a 1963 conference proceeding~\cite{jurney63}, where a  $\sigma_{n\gamma}$($^{207}$Pb) value of 709$\pm$10  mb was deduced from 
(i) a natural Pb cross section of 171$\pm$2 mb~\cite{jowitt59}, (ii)  cross section ratio values  $\sigma_{n\gamma}$($^{204}$Pb)/$\sigma_{n\gamma}$($^{207}$Pb)=0.94$\pm$0.07 
and  $\sigma_{n\gamma}$($^{206}$Pb)/ $\sigma_{n\gamma}$($^{207}$Pb)=0.043$\pm$0.001~\cite{jurney63}.  
The latest 2018 evaluation by S.F.~ Mughabghab~\cite{mughabghab18} gives a $\sigma_{n\gamma}$($^{207}$Pb) value of 647$\pm$9 mb, that is, 9 \% lower, 
since it likely incorporates the results of two experiments that studied the neutron capture of $^{207}$Pb in detail,  
$\sigma_{n\gamma}$($^{207}$Pb)=610$\pm$30 mb from J.C.~Blackmon {\it et al.}~\cite{blackmon02}, and $\sigma_{n\gamma}$($^{207}$Pb)=649$\pm$14 mb from P.~Schillebeeckx {\it et al.}~\cite{schill13}.
It is impossible for us to gauge the quantitative impact of using a larger  $\sigma_{n\gamma}$($^{207}$Pb); 
however, qualitatively this would result in a smaller derived neutron flux, leading to an artificially larger $^{235}$U spectrum,
thus being a possible explanation for the RAA.

An earlier analysis of the nuclear data involved in the ILL data normalization by A.~Letourneau and A.~Onillon  was presented at the 2018 Applied Antineutrino Physics Workshop~\cite{aap2018}.
In this work, they identified discrepancies between the JEFF-3.1~\cite{jeff31} and JEFF-3.3~\cite{jeff33} $\sigma_{n\gamma}$($^{207}$Pb), which can be traced back to the use of the 1981 and 2005 
thermal cross sections evaluations by S.F.~Mughabghab~\cite{mughabghab81, mughabghab05}, respectively.  
They also pointed out an issue with the $^{208}$Pb 7.368 MeV E1 gamma ray K conversion coefficient using the BrIcc code~\cite{bricc}; 
we note that the current BrIcc tables have a 6 MeV upper limit, and our extrapolated value for the 7.368 MeV gamma ray agrees with the one used by ILL.     

In the following section we will briefly discuss the recently published work by Kopeikin {\it et al.}~\cite{Kopeikin21}, which illustrates the normalization issues in the ILL measurements.
This will be followed by a thorough analysis of electron spectra data measured at Oak Ridge National Laboratory in the 1970s, which can provide additional insights into the origin
of the RAA.   Finally, we succinctly discuss possible experimental approaches to measure well normalized electron spectra under equilibrium conditions which would be used to obtain the 
corresponding antineutrino ones.

\section{Kurchatov Institute Measurements}

Kopeikin {\it et al.}~\cite{Kopeikin21} recently published a measurement of the $^{235}$U to $^{239}$Pu electron spectra ratio ($R_{59}$), which is approximately 5\% lower than the one obtained from the ILL measurements.   
In their work, $^{235}$U and $^{239}$Pu foils were placed outside the Kurchatov Institute (KI) reactor core and
electrons were detected using plastic scintillators.   
Succinctly, Kopeikin {\it et al.}~\cite{Kopeikin21} showed that if ILL's $^{235}$U electron spectrum is reduced by 5\%, then a much better agreement with 
Daya Bay's IBD antineutrino yields is achieved, thus possibly eliminating the RAA altogether.    
This groundbreaking work assumes, however, that the $^{239}$Pu and $^{241}$Pu normalizations are paradoxically correct, even though
the $^{235}$U spectrum measured by the same group is not, despite the latter's smaller relative uncertainties and finer energy bin.   
Additionally, both the ILL and KI $R_{59}$ values unexpectedly drop to nearly unity for energies larger than 8 MeV, 
a behavior not relevant for the RAA, but nevertheless disquieting as there is no discernible physical reason for the $^{235}$U  and $^{239}$Pu spectra to have similar values at those energies despite that at lower energies the  $^{235}$U spectrum is considerably larger, 
hence evidencing possible underlying measurement deficiencies. 

\begin{figure}[t] 
\includegraphics[width=0.95\columnwidth, trim=10mm 10mm 25mm 10mm, clip=true]{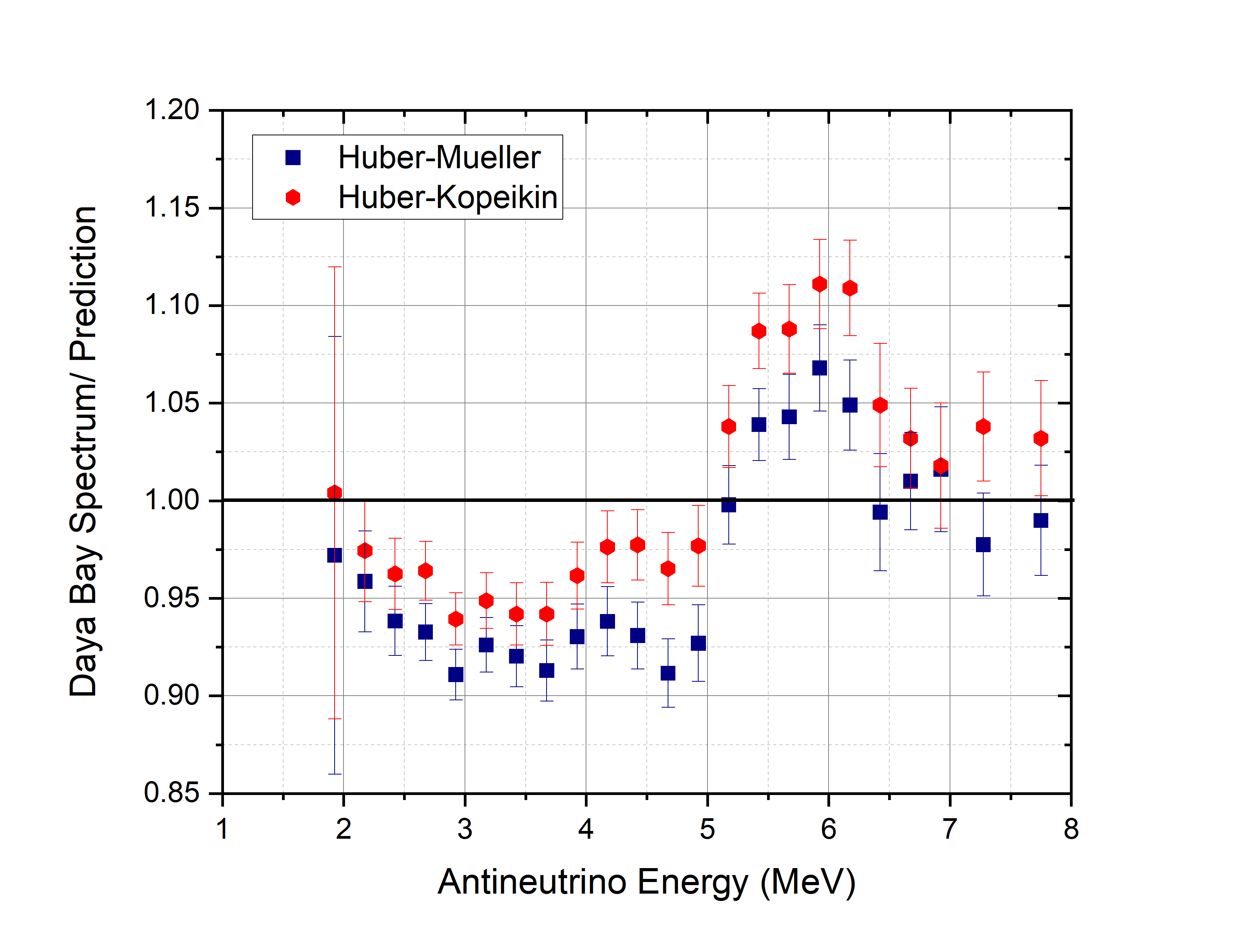}
\caption{
Ratio of the antineutrino spectrum measured by the Daya Bay collaboration to the Huber-Mueller and Huber-Kopeikin models.
}
\label{f.db2models}
\end{figure}

Finally, a simple renormalization of the  $^{235}$U spectrum may not be enough to solve the RAA as can be seen in Fig.~\ref{f.db2models}, 
which shows the ratio of the 2022 Daya Bay antineutrino data~\cite{dayabay22} to the usual Huber-Mueller model as well as what we call the Huber-Kopeikin (HK) model, which uses the $^{239,241}$Pu Huber antineutrino spectra and Kopeikin's $^{235,238}$U ones as listed in the table of Ref.~\cite{Kopeikin21} and assuming a liner interpolation between the energy points.   
As can be seen from this plot, neither the deficit at the IBD spectrum peak, which is the source of the RAA, nor the excess at around 5.5 MeV, known colloquially as 'the bump' are solved with Koepikin's $^{235,238}$U spectra values.

\section{ORNL Measurements}

We have recently encountered electron spectra from experiments performed at Oak Ridge National Laboratory (ORNL) in the 1970s to properly quantify
the decay energy released as function of time following the thermal neutron-induced fission of $^{235}$U~\cite{ornl_235u, ornl_235u_rsicc}, $^{239}$Pu~\cite{ornl_239pu} and  $^{241}$Pu~\cite{ornl_241pu} targets.   
These experiments were performed by a group led by J.~Kirk~Dickens~\cite{Dickens_obit} and were part of a campaign to understand Loss Of Coolant Accident  scenarios~\cite{Dickens_5dh, Dickens_91dh}.   
Target foils were irradiated inside the Oak Ridge Research Reactor and later placed in front of a detection system using a rabbit mechanism.   
Gamma and electron spectra were measured using scintillator detectors, which were normalized 
per fission by using well-known fission products' gamma decay intensities.
Electron spectra for $^{20}$F and $^{56}$Mn were measured separately, which agreed quite well with calculated ones using nuclear databases, adding an important element of confidence to their experimental results. 
The electron data for energies above 1.5~MeV are well accounted for by summation calculations; for lower energies, the electron spectra contain contributions from gamma rays and conversion electrons.

\begin{figure}[t] 
\includegraphics[width=0.95\columnwidth, trim=10mm 10mm 25mm 10mm, clip=true]{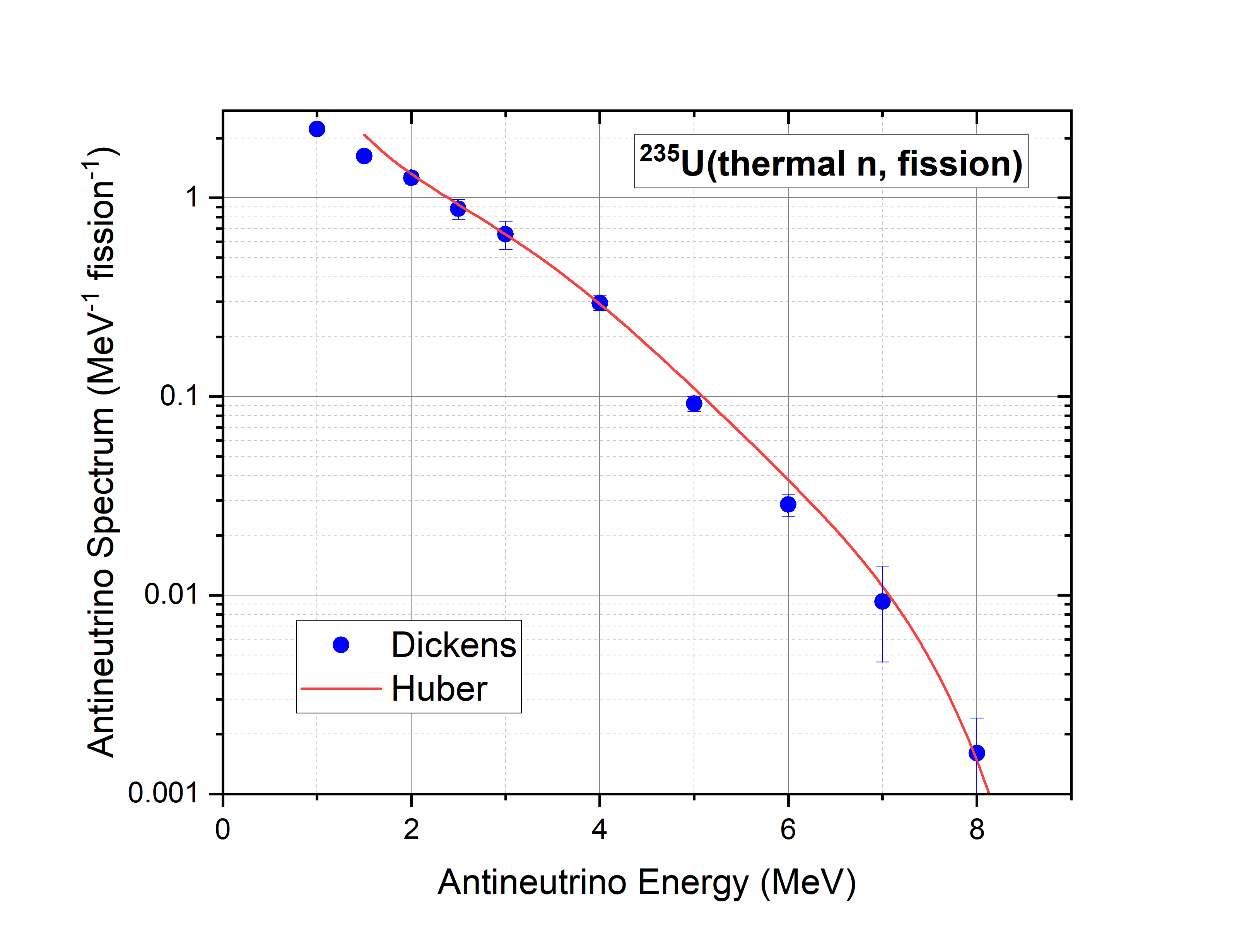}
\caption{
Comparison between the Dickens and Huber antineutrino spectrum  following the thermal-neutron induced fission of  $^{235}$U.   
}
\label{f.5udickensan}
\end{figure}

The data  in these reports correspond to three irradiations:  a short 1-second one, a medium one of 5 or 10 s, and a long one of 50 or 100 s.   The time interval between the irradiation and start of counting, $T_{cs}$, was 1.7 s for the short irradiation, 10.7 or 17.7 s for the medium ones, and  170 or 250 s for the long ones.  Data were counted for 110 to 130 s, 795 to 1,198  s, and 13,500 to 14,000 s for the three irradiations, respectively.   
The reports contain 13 to 15 spectra per irradiation, corresponding to an increasing counting interval to maintain an approximately similar number of counts in the lower energy portion of the spectra.  
These reports were submitted to the  Nuclear Science References database~\cite{nsr}, and  the data  were submitted to the EXFOR database~\cite{exfor} following their digitization.
We note that the $^{235}$U  and $^{239}$Pu beta decay heat values derived from the ORNL data agree well with those measured by Akiyama and San~\cite{akiyama82,hagura06};
and interestingly, the electron data for $^{235}$U, obtained by adding up all the individual electron spectra for the short irradiation, were used to obtain the corresponding equilibrium antineutrino spectrum in 1981~\cite{Dickens81}, a work which despite its pioneering relevance has not been cited by the many reactor antineutrino articles published in the last 10 years, and which agrees surprisingly well with the Huber values, as shown in Fig.~\ref{f.5udickensan}, despite the 30-year time lapse between them.  

Due to the relative similarities between the ORNL experimental setup and Kopeikin's, our goal 
would be to determine if we can obtain equilibrium spectra ratios from  the ORNL data to elucidate possible normalization issues in the ILL electron spectra, 
despite that none of the irradiation conditions correspond to an equilibrium situation.

\section{Formalism}

\begin{figure}[t] 
\includegraphics[width=0.95\columnwidth, trim=20mm 10mm 25mm 10mm, clip=true]{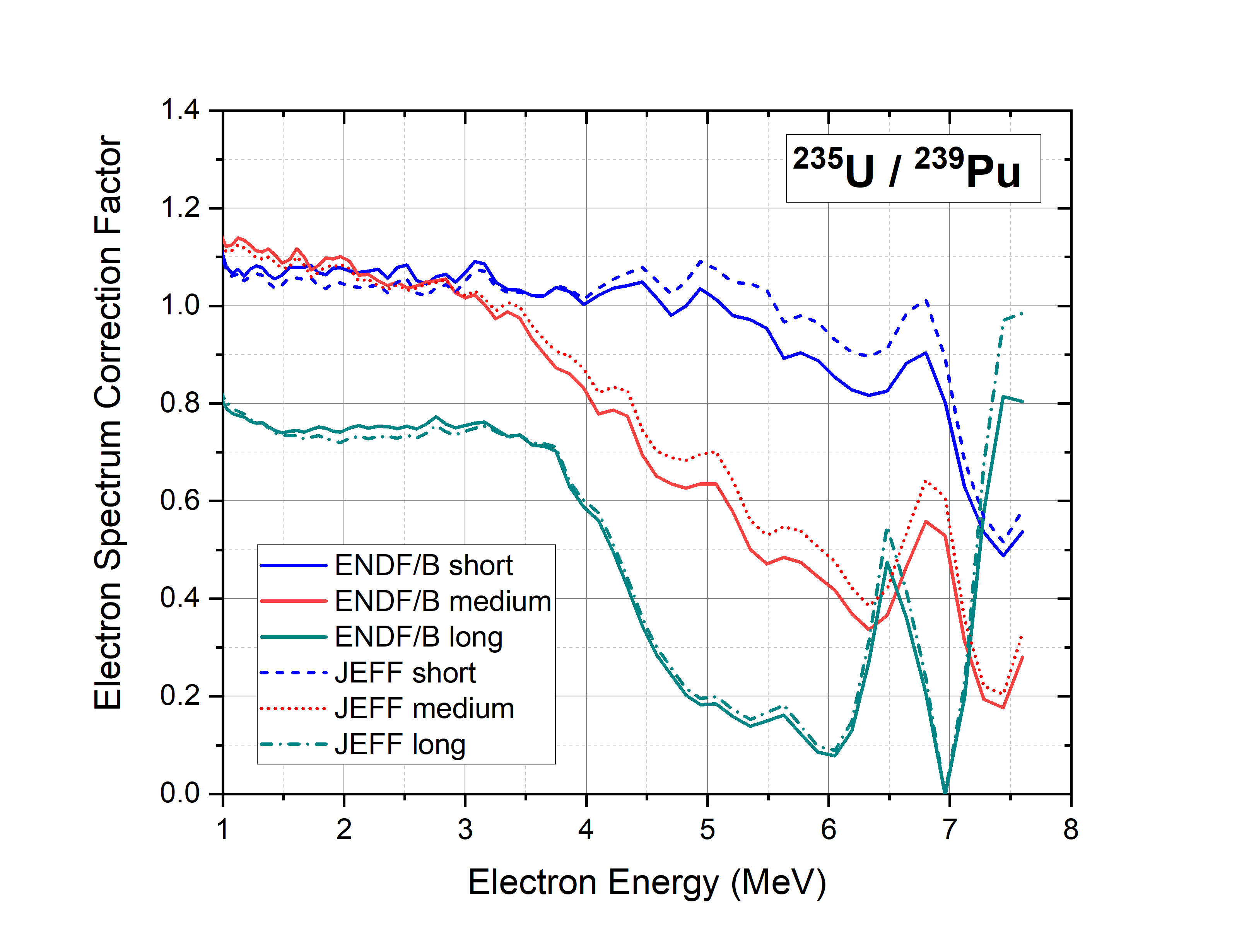}
\caption{
Correction term needed to convert the measured ratio of $^{235}$U to $^{239}$Pu electron spectra to the corresponding spectra ratio under equilibrium conditions.
}
\label{f.c5to9}
\end{figure}

Using the summation method~\cite{vogel81, sonzogni15,estienne19, schmidt21}, the electron and antineutrino spectra  from a target $a$ under equilibrium equilibrium conditions can be calculated as 
\begin{equation}
S^a_{s,eq}=\sum CFY^a_j S_j,  
\end{equation}
where $CFY^a_j$ are the cumulative fission yields and $S_j$ the  corresponding electron or antineutrino spectra.   
Due to the linear dependence on $CFY$s, 
we can convert a spectrum measured with irradiation conditions labeled with the index $i$, $S^a_{m,i}$, to the corresponding equilibrium spectrum as
\begin{equation}
S^a_{eq,i}=S^a_{m,i} + \sum (CFY^a_j - Y^a_{j,i} ) S_j,
\label{eq.s1eqi}
\end{equation}
where $Y^a_{j,i}$ are the effective cumulative yields during the experiment. 
Assuming that the nuclear data in the network, such as fission yields, decay branching ratios, half-lives and electron/antineutrino spectra  are of high fidelity,
we can replace the $ \sum (CFY^a_j - Y^a_{j,i} ) S_j$  term by $(S^a_{s,eq}-S^a_{s,i})$, where $S^a_{s,i}$
is the summation spectrum calculated with the $i$-irradiation conditions.
Therefore, the ratio between two spectra obtained this way would be 
\begin{equation}
R_{ab,i,k}=\frac{S^a_{eq,i}}{S^b_{eq,k}}=\frac{S^a_{m,i} +S^a_{s,eq}-S^a_{s,i}}{S^b_{m,k}+S^b_{s,eq}-S^b_{s,k} },
\label{rab}
\end{equation}
which can be written as
\begin{equation}
R_{ab,i,k}=\frac{S^a_{m,i}}{S^b_{m,k} }C_{ab,i,k},
\end{equation}
with
\begin{equation}
C_{ab,i,k}=\frac{1 +(S^a_{s,eq}-S^a_{s,i})/S^a_{m,i}}{1+(S^b_{s,eq}-S^b_{s,k})/S^b_{m,k} },
\end{equation}
factored out to help us understand how different $R_{ab,i,k}$ would be from the ratio of measured spectra $S^a_{m,i}/S^b_{m,k}$.

\begin{figure}[t] 
\includegraphics[width=0.95\columnwidth, trim=20mm 10mm 25mm 10mm, clip=true]{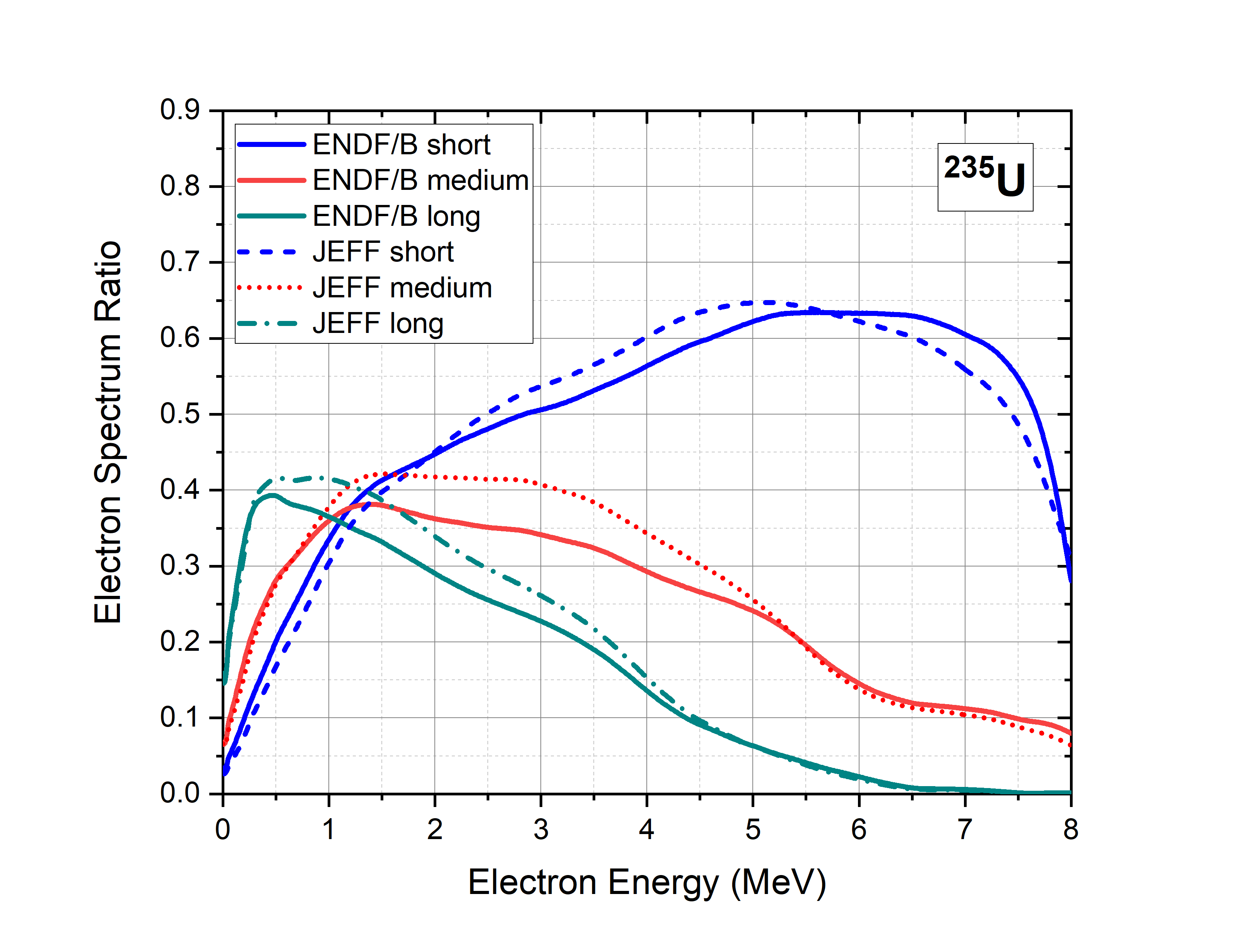}
\caption{
Summation electron spectrum  following the thermal-neutron induced fission of  $^{235}$U under three different irradiation conditions, divided by the corresponding summation equilibrium spectrum.   
}
\label{f.s2e5}
\end{figure}

Plots of the $C_{ab,i,k}$ term for the $^{235}$U to  $^{239}$Pu equilibrium electron spectra ratio are given in Fig.~\ref{f.c5to9} for the three
ORNL irradiation conditions.  We have used the  fission yields from the JEFF-3.3 library~\cite{jeff33}, and alternatively, the decay data from JEFF-3.3 or an updated version of the
 ENDF/B-VIII.0~\cite{endfb8} one.   
For the short irradiation and energies lower than 5~MeV, this term is very close to unity and the differences between the two decay data libraries are minimal;
at higher energies, fluctuations in $C_{ab,i,k}$ are due to statistical effects in the measured spectra.   
For the medium and long irradiations, correction factors are more important and reliance on the summation method is higher.
We can understand this fact from Fig.~\ref{f.s2e5} which shows $S^a_{s,i}/S^a_{s,eq}$ plots for the three $^{235}$U irradiations.
We can see that in the 1.5 to 7 MeV interval, the short irradiation  accounts for 40-65\% of the equilibrium spectrum, 
while the medium and long irradiations account for significantly less because of their considerably larger $T_{cs}$ value.     

\begin{figure}[t] 
\includegraphics[width=0.95\columnwidth, trim=2mm 2mm 2mm 2mm, clip=true]{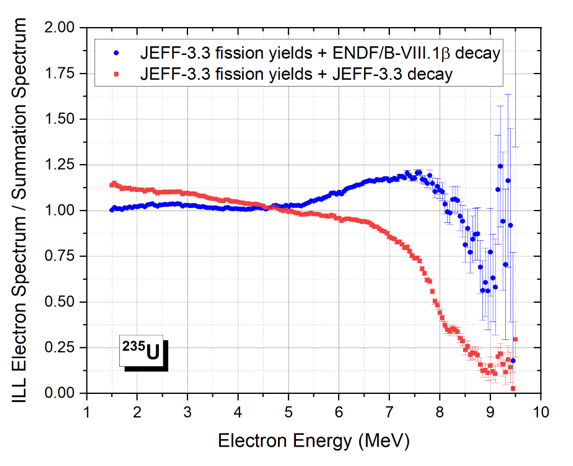}
\caption{
Ratio of the $^{235}$U electron spectrum measured at ILL to calculations using the JEFF-3.3 fission yields and alternatively the 
JEFF-3.3 or updated ENDF/B-VIII.0 decay data.  Uncertainties are those from the ILL spectrum only.}
\label{f.5ill2db}
\end{figure}

The differences observed when using the JEFF-3.3 or ENDF/B-VIII.0 decay data are mainly due to 
the implementation of Total Absorption Gamma Spectroscopy  beta  intensities in the latter, see for instance Refs.~\cite{greenwood97, algora10, rasco16}.   This topic is explored
in more detail in Fig.~\ref{f.5ill2db}, which plots the $^{235}$U electron spectrum measured at ILL divided by calculations that employ the
JEFF-3.3 cumulative fission yields and alternatively the JEFF-3.3 or updated ENDF/B-VIII.0 decay data.   
For energies less than 5 MeV, the ILL to JEFF decay ratio is larger than 1 due to the lack of TAGS beta intensities in it,
while for energies higher than 6 MeV the ratio is considerably lower than one since the JEFF-3.3 decay data sub-library doesn't contain theoretical electron
spectra for fission products with incomplete decay data~\cite{cgm}. 
Similar results are observed for $^{239}$Pu and $^{241}$Pu.

Moving forward with our analysis, because of its small correction, the short irradiation results will be considered the most reliable of the three, 
and the updated ENDF/B-VIII.0 decay data will be used due to its higher fidelity in this particular application. 

The  $Y^a_{j,i}$ terms in Eq.~\ref{eq.s1eqi} were obtained by numerically solving the corresponding Bateman's equations, which during the irradiation are
\begin{equation}
\frac{dN_j(t)}{dt} = R_f IFY_j - \lambda_j N_j + \sum b_{jk} \lambda_k N_k,
\end{equation}
where $R_f$ is the fission rate, while  $IFY_j$, $\lambda_j$, and  $N_j$ are the independent fission yields, decay constant and population for the j-th fission product,
and $b_{jk}$ is the nuclear decay probability from the k-th to the j-th element in the decay network.   After the irradiation, the term  $ R_f IFY_j$ disappears from this equation.
Finally, $Y^a_{j,i}$ is obtained by integrating $N_j(t)$ during a specific counting interval.
   
\begin{equation}
Y^a_{j,i} = N^{-1}_f\int N_j(t) dt,
\end{equation}
where $N_f=\int R_f dt$,  is the number of fission events during the irradiation.

\begin{figure}[t] 
\includegraphics[width=0.95\columnwidth, trim=2mm 2mm 6mm 2mm, clip=true]{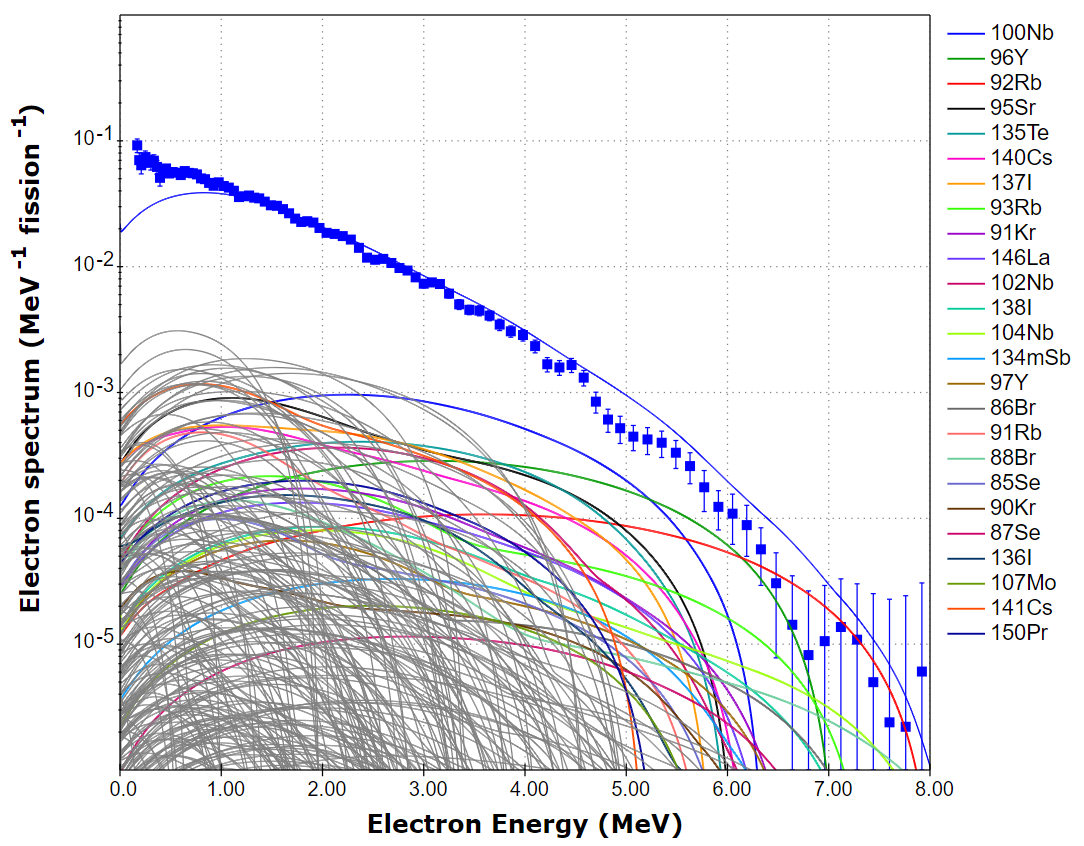}
\caption{
Experimental and calculated $^{239}$Pu electron spectra for the short irradiation, with a  waiting time of 19.7 seconds and a counting time of 5 seconds.   
The contributions of the individual fission products are also plotted, and in particular, the largest contributors are 5 MeV are highlighted in color.
}
\label{f.9spaghetti}
\end{figure}

As an example, the ORNL electron spectrum for the short irradiation on $^{239}$Pu, with a waiting time of 19.7 seconds and a counting time of 5 seconds is compared with the corresponding 
$\sum Y^a_{j,i} S_j$ term in Fig.~\ref{f.9spaghetti},
highlighting also the most important contributors at 5 MeV of electron energy.   
Overall, the agreement between the ORNL electron data and the summation calculations is  good; however, summation calculations tend to overestimate the sum spectra at the higher energies,
with differences of up to 20\%.   We note that this disagreement has a negligible impact in the conclusions about the ILL normalizations drawn later in this article.

\begin{figure}[t] 
\includegraphics[width=0.95\columnwidth, trim=1mm 1mm 1mm 1mm, clip=true]{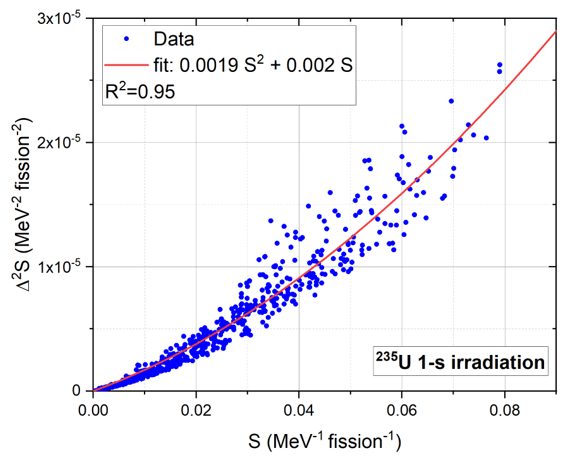}
\caption{
Square of the electron spectrum uncertainty as function of electron spectrum for the short $^{235}$U irradiation, including 
a quadratic fit to the data to obtain the uncertainty of the sum spectrum.
}
\label{f.5u_1s_unc}
\end{figure}

 Uncertainties in the $R_{ab,i,k}$ term will have experimental  and databases contributions obtained as described below.
\begin{itemize}
\item Uncertainties in the $(S^a_{s,eq}-S^a_{s,i})$ terms are calculated using a Monte Carlo (MC) method, where
 for each history the  independent fission yields, half-lives, decay branching ratios and electron spectra were varied, leading to a new set of cumulative fission yields, and a new spectrum $S^a_{s,i}$ after solving the corresponding Bateman's equations;
 this approach validity was confirmed by calculating delayed neutron activities~\cite{bdncrp} mean values and standard deviations, which agreed well with evaluated ones~\cite{jeff33}.
\item Since the experimental uncertainties $\Delta S^a_{m,i}$ are not available we derived them from the uncertainties
of the individual spectra that are summed to obtain $S^a_{m, i}$, 
assuming that for electron energies larger than 1 MeV, spectra uncertainties are the sum of a statistical and a systematic term, 
$\Delta^2 S(E) = \Delta^2 S_{stat}(E) +\Delta^2 S_{sys}(E)$, 
with the former proportional to the square root of the spectrum, $\Delta S_{stat}(E) = c_{stat} S^{1/2}(E)$, and the latter proportional to the spectrum, $\Delta S_{sys}(E)=c_{sys} S(E)$.
The $c_{stat}$ and $c_{sys}$ parameters were obtained from a fit to the approximately 700 to 900 ($S(E)$, $\Delta S(E)$) pairs of points per irradiation per target,
with coefficients of determination $R^2$ values in the 0.74 to 0.98 range, noting that the long irradiation $^{241}$Pu data accounts for the lower $R^2$ values; those $c_{stat}$ and $c_{sys}$ parameters were later used to
obtain the $\Delta S^a_{m, i}$ values.  
As an example, Fig.~\ref{f.5u_1s_unc} shows the square of the electron spectrum uncertainties as function of spectrum values for the short $^{235}$U irradiation, including the quadratic fit used to obtain
the uncertainties in the sum spectrum.
\item Experimental $\Delta S^a_{m,i}$ and MC $\Delta (S^a_{s,eq}-S^a_{s,i})$ uncertainties were added in quadrature assuming no correlation between them.
\item $\Delta R_{ab,i,k}$ were obtained employing a first order Taylor expansion assuming no correlations between Eq.~\ref{rab}'s numerator and denominator.
\end{itemize}
 
\begin{figure}[t] 
\includegraphics[width=0.95\columnwidth, trim=20mm 10mm 25mm 10mm, clip=true]{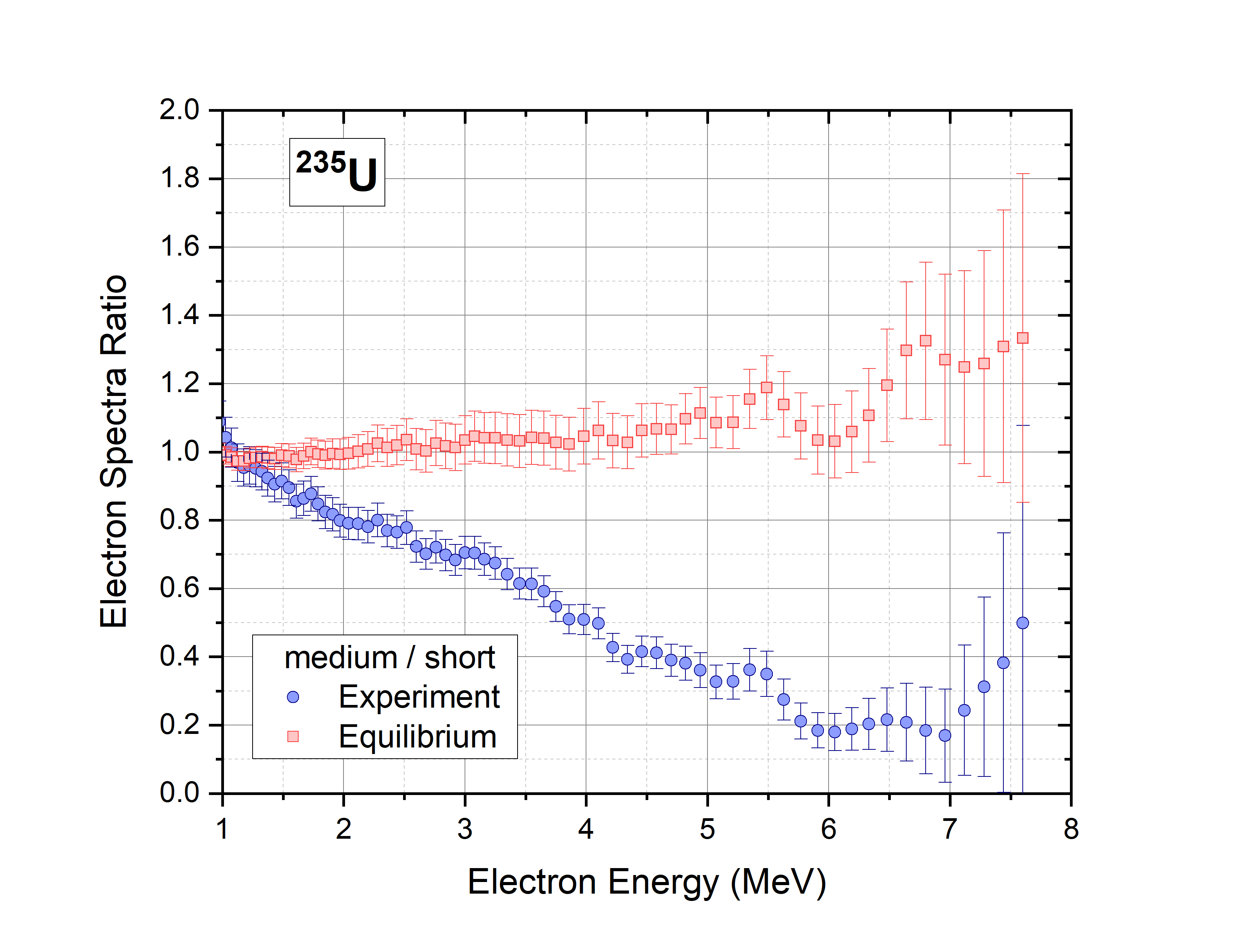}
\caption{
Ratio of the experimental and equilibrium $^{235}$U medium irradiation electron spectrum to the corresponding short one.
}
\label{f.r55_10_1}
\end{figure}

An interesting feature about Eq.~\ref{rab} is that $R_{aa}$ values for the three different possible irradiation combinations would provide a consistency check 
since they should be equal to unity.  
This is shown for the ratio of $^{235}$U's medium to short irradiation spectra in Fig.~\ref{f.r55_10_1}, with similar results for the other target and irradiation combinations.
Consequently, we conclude that the most reliable energy range for our results is 1.5 to 5~MeV; 
for higher energies, uncertainties due to diminished statistics become dominant, 
not surprising in an experimental campaign designed to obtain $\beta$ decay heat values,  which are proportional to $\int E S(E) dE$ with the integrand peaking in the 2 - 3 MeV region, 
rather than highly precise $S(E)$ values for energy values higher than 5 MeV.

\section{Results}

\begin{figure}[t] 
\includegraphics[width=0.95\columnwidth, trim=20mm 10mm 25mm 10mm, clip=true]{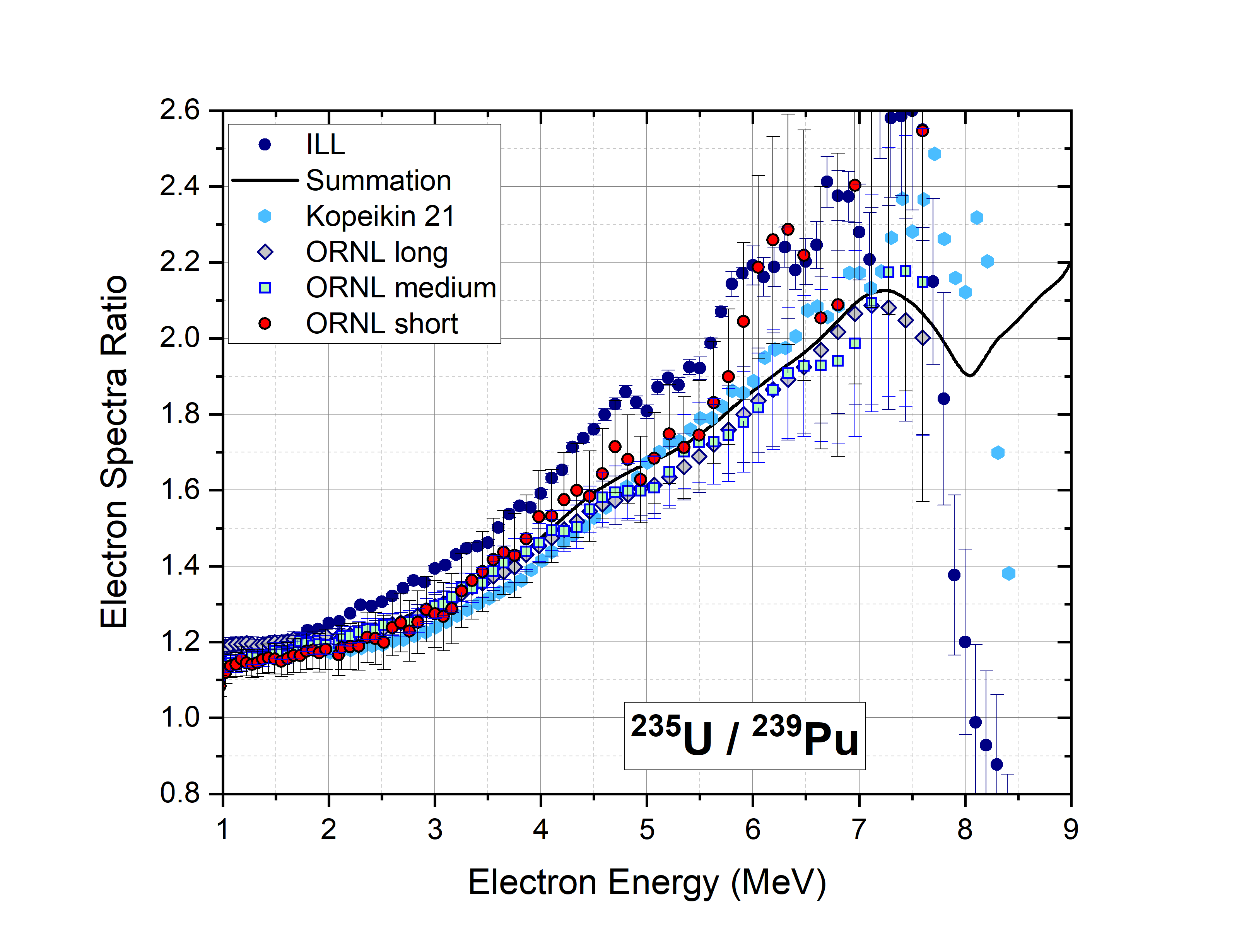}
\caption{
Ratio of $^{235}$U to $^{239}$Pu electron spectra for thermal neutron-induced fission in equilibrium conditions.
}
\label{f.5to9}
\end{figure}

\begin{figure}[t] 
\includegraphics[width=0.95\columnwidth, trim=20mm 10mm 25mm 10mm, clip=true]{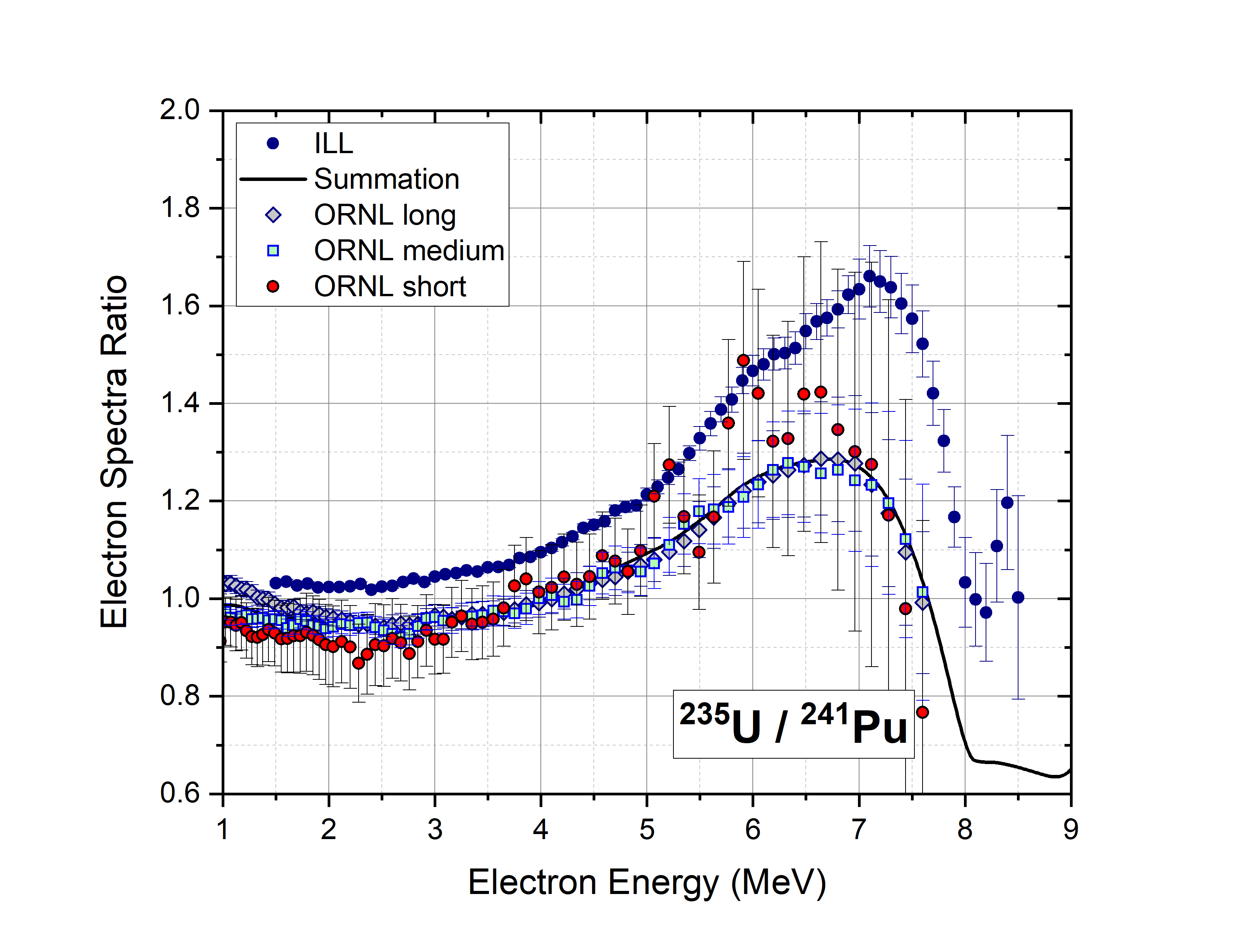}
\caption{
Ratio of $^{235}$U to $^{241}$Pu electron spectra for thermal neutron-induced fission in equilibrium conditions.
}
\label{f.5to1}
\end{figure}

\begin{figure}[t] 
\includegraphics[width=0.95\columnwidth, trim=20mm 10mm 25mm 10mm, clip=true]{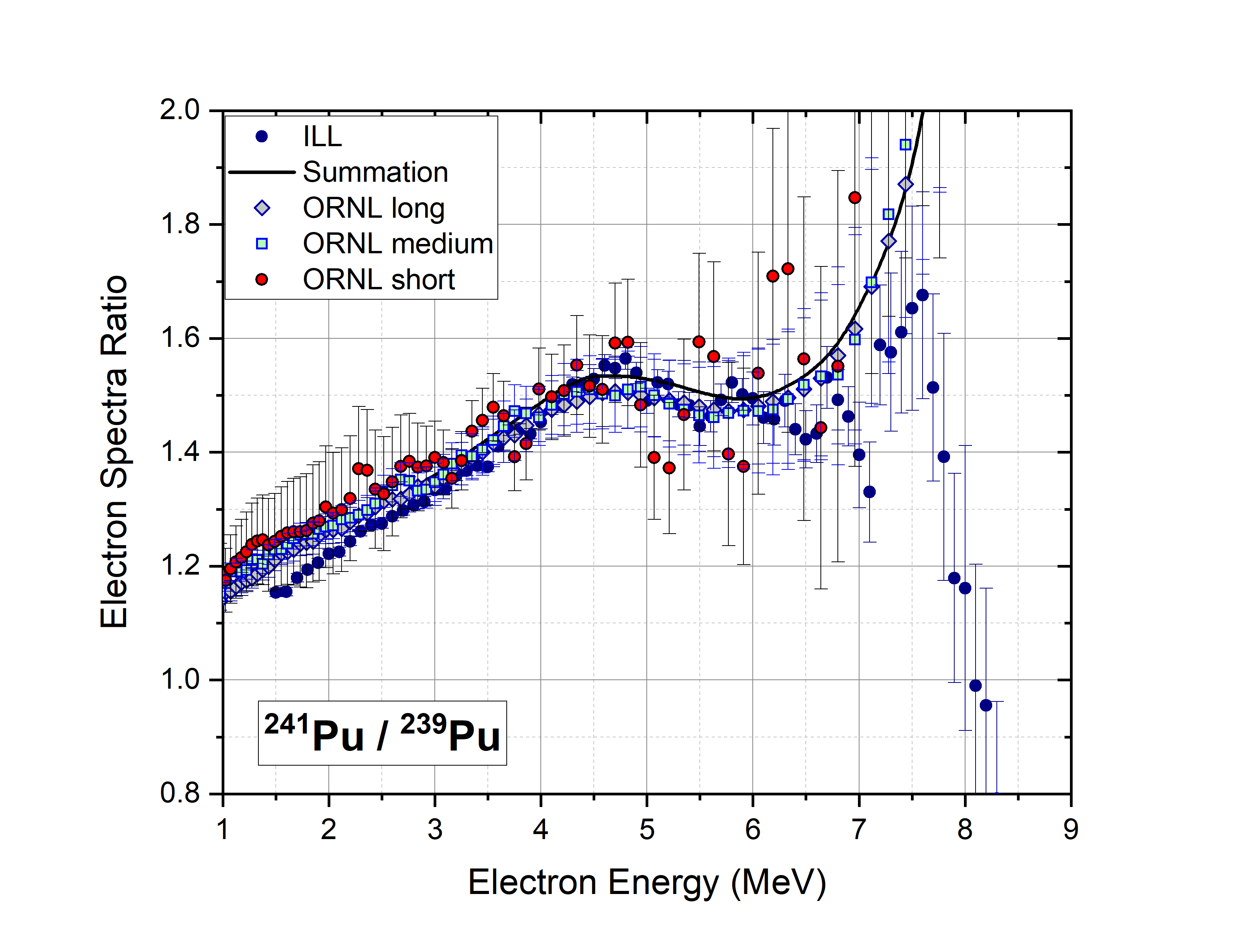}
\caption{
Ratio of $^{241}$Pu to $^{239}$Pu electron spectra for thermal neutron-induced fission in equilibrium conditions.
}
\label{f.1to9}
\end{figure}

Results for the $^{235}$U to  $^{239}$Pu electron spectra ratio ($R_{59}$) are given in Fig.~\ref{f.5to9},
compared to the ILL and summation values, as well as the KI ones, the latter of which were read off the plot as the data were not made available.    
Fig.~\ref{f.5to1} shows the  $^{235}$U to  $^{241}$Pu spectra ratio  ($R_{51}$) results compared to the ILL and summation values, 
while results for the  $^{241}$Pu to  $^{239}$Pu spectra ratio  ($R_{19}$) can be seen in Fig.~\ref{f.1to9}.   
As mentioned earlier, our most reliable results are those from the short irradiation in the 1.5 to 5 MeV energy interval;
despite their larger correction,  results from the medium and long irradiation are also shown, which will track closely with the summation results, particularly for energies larger than
3.5 MeV, where the correction factors are more dominant.

Overall, we observe differences with the ILL ratio values, 
and in particular (i)  our $R_{59}$ values align with the lower trend observed by Kopeikin {\it et al.}~\cite{Kopeikin21}; however,
at around 3.5-4.5 MeV, where short irradiation corrections are minimal, our results are approximately half-way between the ILL and KI measurements.   
(ii) The $^{241}$Pu electron spectrum seems larger than the ILL and summation results as evidenced by smaller $R_{51}$ and larger $R_{19}$
values.    
(iii)  The ILL $R_{51}$ values follow the summation calculations trend for higher energies, but 
the ILL $R_{59}$ and $R_{19}$ as well as KI $R_{59}$ ones do not, which could indicate issues with the $^{239}$Pu data at those energies;
without attempting to provide an explanation, we nevertheless note that
a small presence of $^{241}$Pu or $^{235}$U in the $^{239}$Pu target could cause that behavior.
Finally, we think that at this stage it is unlikely that we could use the ORNL data to normalize the ILL electron spectra to obtain new
antineutrino spectra due to the lack of the ORNL sum spectra uncertainties and correlations. 

Because of the lack of agreement between the $R_{59}$, $R_{51}$ and $R_{19}$ values as shown in Figs.~\ref{f.5to9} to \ref{f.1to9},
it is clear that the best way forward would be to remeasure the electron spectra using an experimental technique employing  the best features of the ILL, KI and ORNL experiments,
while minimizing nuclear databases input.
For precise normalization, counting of foils placed outside the reactor core would be needed, while for the measurement of the
electron spectra, using detectors with superconducting solenoids, such as the one in Ref.~\cite{solenoid}, would be desirable since they would block gamma rays
and conversion electrons while providing superb energy resolution;
in particular, the use of at least two of these detectors, with one counting at a fixed electron energy value, would define the spectrum's energy dependence
even better.   
Additionally, a measurement of the $^{238}$U electron spectrum at fast neutron energies using the same setup would be needed as the nuclear data behind the corresponding Mueller model~\cite{mueller11} has been considerably improved and the only extant $^{238}$U measurement~\cite{haag14} requires knowledge of the $^{235}$U electron spectrum to deduce it.
This experimental campaign would also yield precise values of the energies carried away by the antineutrinos,  $\langle E_{\nu}\rangle$,  to calculate the total energy released following fission~\cite{kopeikin04, ma13},
improve decay heat data~\cite{nichols23}, as well as help benchmark fission yield and nuclear decay databases.

\section{Conclusions}

In summary, we have deduced the electron spectra ratio under equilibrium conditions  $R_{59}$, $R_{51}$ and $R_{19}$ 
from the electron spectra measured by Dickens~{\it et al.} with the assistance of summation calculations that 
employ the latest nuclear databases.
Our most reliable $R_{59}$ values differ from the ILL ones and are in better agreement with those reported
by Kopeikin {\it et al.}, supporting the hypothesis that the Reactor Antineutrino Anomaly may be mainly due to
faulty $^{235}$U electron spectrum normalization.   
Additionally, our $R_{19}$ values are also higher than those from ILL,
indicating that the normalization for $^{239}$Pu and $^{241}$Pu may not be as precise as needed.
These conclusions are supported by our survey of the $^{207}$Pb thermal neutron cross section data, with a recommended value lower than the one used at ILL, 
and affecting the $^{235}$U and $^{241}$Pu ILL normalizations, 
as well as our assessment of the $R_{59}$ behavior at high energies indicating a possible contamination in the $^{239}$Pu target.
As a consequence, we think that a new experimental campaign to measure electron spectra at a location that allows a precise
normalization and using a spectrograph that provides high energy resolution and signal-to-noise ratio is needed
to finally understand the electron antineutrino spectrum produced by nuclear reactors.

\begin{acknowledgements} 
Work at Brookhaven National Laboratory was sponsored by the Office of Nuclear Physics, Office of Science of the U.S. 
Department of Energy under Contract No. DE-AC02-98CH10886,
as well as by the U.S. Department of Energy, 
National Nuclear Security Administration, Office of Defense Nuclear Nonproliferation Research and Development (DNN R\&D).
We are grateful to Catherine Dunn for assistance in digitizing the ORNL reports tables.
We acknowledge fruitful discussions with Krzysztof Rykaczewski, Charlie Rasco, Robert Grzywacz, Mitch Allmond and Lowell Crow (ORNL), 
as well as Tibor Kibedi and Andrew Stuchbery (ANU) in preparation of the WoNDRAM workshop~\cite{wondram}, which helped delineate our future experimental activities statements.
We are also grateful to David Glasgow (ORNL) for guidance in trying to recover the full ORNL electron spectra data.
Finally, our work on the $^{207}$Pb(n,$\gamma$) cross section was motivated by discussions at the 
2$^{nd}$ IAEA Technical Meeting on Nuclear Data Needs for Antineutrino Spectra Applications, organized
by P. Dimitriou.

\end{acknowledgements}

\end{document}